\documentclass[12pt]{article}

\usepackage{amsmath,amssymb,amsthm,mathtools}
\usepackage{graphicx}
\usepackage{natbib}
\usepackage{url}
\usepackage{enumitem}
\usepackage{booktabs}
\usepackage{multirow}
\usepackage{array}
\usepackage{algorithm}
\usepackage{algpseudocode}
\usepackage{float}
\usepackage{color}
\usepackage[colorlinks=true,linkcolor=blue,citecolor=blue,urlcolor=blue,anchorcolor=blue]{hyperref}

\newcommand{\blind}{0}

\def\I{\mathbb{I}}
\def\P{\mathbb{P}}
\def\E{\mathbb{E}}
\def\R{\mathbb{R}}
\def\trans{^{\scriptscriptstyle\sf T}}
\newcommand{\rh}{\widehat r}
\newcommand{\rF}{r_F}
\newcommand{\Chat}{\widehat C}
\newcommand{\X}{\mathcal X}
\newcommand{\pkgname}{\texttt{depthinfer}} 
\DeclareMathOperator{\Var}{Var}
\DeclareMathOperator*{\argmax}{argmax}

\newtheorem{assumption}{Assumption}

\newtheorem{theorem}{Theorem}
\newtheorem{lemma}{Lemma}

\theoremstyle{remark}
\newtheorem{remark}{Remark}

\begin{document}

\def\spacingset#1{\renewcommand{\baselinestretch}{#1}\small\normalsize}\spacingset{1}

\if0\blind
{
  \title{\bf Simultaneous Inference for Centre-Outward Rankings of Random Objects}
  \author{Lingfeng Lyu\textsuperscript{1,2} and Doudou Zhou\textsuperscript{1}\thanks{Corresponding author: ddzhou@nus.edu.sg}\bigskip\\
    \small \textsuperscript{1} Department of Statistics and Data Science, National University of Singapore\\
    \small \textsuperscript{2} Department of Statistics and Finance, University of Science and Technology of China
   }  
  \date{}
  \maketitle
} \fi

\if1\blind
{
  \bigskip\bigskip\bigskip
  \begin{center}
    {\LARGE\bf Simultaneous Inference for Centre-Outward Rankings of Random Objects}
  \end{center}
  \medskip
} \fi

\bigskip
\begin{abstract}
In many applications one object is chosen to represent a sample and the least typical objects are set aside, usually without any measure of uncertainty. We measure how central an object is by its reference depth, the probability that a random reference point is closer to it than to a random opponent. Using only pairwise dissimilarities, we construct simultaneous intervals for all ranks, a centre set that contains the most central observation under stated conditions, and outlier sets that contain only objects among the least central ones, each with asymptotic probability at least a prescribed level. These guarantees hold even though the ranked objects are themselves random. For symmetric distributions on the real line, the centre set has the smallest possible order of size up to a logarithmic factor, and in dimension three or more it reduces to the most central observation with probability tending to one. For typical meteorological years the centre set usually contains several candidate years, so the data do not determine the year on which building-energy simulations rest; for three collections of gene trees it contains one tree, whose features can be reported as those of the most central tree.
\end{abstract}


\vfill
\newpage
\spacingset{1.45}

\section{Introduction}\label{sec:intro}

Building-energy simulations are run on a typical meteorological year (TMY), which joins twelve months of recorded weather, one for each calendar month at a weather station. For each station and calendar month, the TMY3 procedure of the US National Renewable Energy Laboratory \citep{wilcox2008}, a revision of the method of \citet{hall1978tmy}, ranks that month of each past year, called a candidate year, by how close its daily weather is to the long-term distribution, keeps the five best as a shortlist and selects one of them, which every simulation for that station then uses. The procedure does not report how certain this choice is. With only 40 to 50 candidate years, can the data tell which candidate year is the most typical? We measure how typical a candidate year is by how central it is among the others, and construct a 95\% confidence set for the most central candidate year. At Olympia Airport, Washington, where 49 Januaries of 1973--2022 have enough valid days, this set contains nine Januaries, including 2004, which TMY3 ranks first, and 2002, the most central January by our estimate. Choosing among these nine changes the heating degree days, which drive the energy used for heating, by up to 15\% of their mean over the 49 Januaries (Supplementary Section~S6.1). Across the 433 combinations of station and calendar month analysed in Section~\ref{sec:tmy}, the set has more than five members in 85\% of the combinations, and in about half of them the TMY3 shortlist includes a candidate year that the data rule out as the most central. Energy estimates made with a TMY, and the design decisions based on them, therefore rest on a choice that the data often cannot settle.

The same problem arises whenever one object is chosen to represent many, or the least typical objects are set aside. Phylogenomic studies estimate hundreds or thousands of gene trees and report a central one as a summary \citep[e.g.][]{nye2017pca}. Candidate years and gene trees, like functions, networks or points in a learned embedding, have no natural order, and often the only information available is a dissimilarity between each pair. Statistical depth orders such data from the centre outwards \citep{Tukey1975,Liu1990,zuo2000general}, and recent proposals define depth, quantiles and ranks from distances alone \citep{liu2022quantiles,wang2024mdf,geenens2023statistical,dai2023tukey,virta2023}. These constructions give a population ordering and a consistent estimate of it. In practice the most central object and the outliers are then chosen by convention, such as keeping a fixed number of the highest-ranked candidates, without a measure of the uncertainty in the ordering.

In this paper we study which observed objects could be the most central one and which are outlying. We focus on the reference depth $r_F(y)$ of an object $y$, the probability that a random reference point is closer to $y$ than to a random opponent, where the reference point and the opponent are independent draws from the population distribution $F$. For a chosen level $\alpha$, we construct a centre set that contains the most central observation and, for each $k$, an outlier set that contains only observations among the $k$ least central. Each of the two statements holds with asymptotic probability at least $1-\alpha$, the second for all $k$ at once. We also construct simultaneous intervals for all ranks. Here the rank of an observation is its position when the observations are ordered by decreasing value of $r_F$.

Three features make the problem different from standard ranking inference. First, the targets are indexed by the sample: the 49 Januaries at Olympia are both the data and the objects to be ranked. We want to know which of the observations $X_1,\dots,X_n$, an independent sample from $F$, is most central, so the parameters $r_F(X_1),\dots,r_F(X_n)$ are random and are estimated from the same observations that index them. Second, the estimated depths of different observations share their reference points and opponents, so they are strongly dependent, almost perfectly so for two nearby objects. Third, the precision of the estimated depths varies greatly across the sample. The depths of the most central observations are estimated much more precisely than those of observations at moderate distance from the centre of the distribution, so intervals of a common width would be too wide for the most central observations.

Our first contribution is a simultaneous confidence band for the depths of the observed objects: intervals for $r_F(X_1),\dots,r_F(X_n)$ that contain all $n$ depths at once with asymptotic probability at least $1-\alpha$. We choose the width of the band with a studentised bootstrap. This coverage holds although the targets are indexed by the sample, because as the sample grows the observations cover the support densely, so the maximum standardised error over the observed objects behaves like a maximum over the whole support (Theorems~\ref{thm:full} and \ref{thm:guard}).

Our second contribution is to build the centre set and the outlier sets from pairwise comparisons of estimated depths, which exploit the correlation between estimated depths. For each observation separately, the centre set tests whether it could be the most central one, using only the other observations whose estimated depths are close to or above its own \citep{andrewssoares2010}. Testing each candidate separately makes the centre set smaller than a set that uses one critical value for all comparisons. Under regularity conditions, we prove coverage of the centre set for Euclidean data, for metric spaces of high intrinsic dimension and, under further local conditions, for metric spaces of low intrinsic dimension.

Third, we study how small a centre set can be. On the line, the centre set has size of order $(n\log n)^{1/2}$ for symmetric distributions and at most of order $n^{2/3}(\log n)^{1/3}$ in general. Since every procedure whose coverage holds uniformly over local shifts in a location family has expected size at least of order $\sqrt n$, the first rate is optimal up to a factor $(\log n)^{1/2}$ (Theorems~\ref{thm:row1d} and \ref{thm:1dlower}). In the plane the centre set grows at most polylogarithmically, and from dimension three onwards it consists of the most central observation alone with probability tending to one.

\subsection{Related work}\label{sec:related}

Classical approaches to ranks and centrality beyond the real line include componentwise and spatial ranks \citep{puri1966class,hallin1995multivariate,chaudhuri1996geometric,oja2010multivariate}, depth-induced orderings \citep{liu1993quality,zuo2000general,Zuo2003,Yeon2025JRSSB}, Mahalanobis and metric-based constructions \citep{hallin2002optimal,pan2018ball}, and more recent transport- or graph-based formulations \citep{deb2021multivariate,zhou2023new}. In metric spaces, the distance-based constructions include the metric ranks and quantiles of \citet{liu2022quantiles}, the metric distribution function of \citet{wang2024mdf}, metric lens depth \citep{liu2011lens,kleindessner2017lensdepth,geenens2023statistical}, Tukey depth for object data \citep{dai2023tukey} and metric spatial depth \citep{virta2023}. \citet{dubey2024metric} and \citet{zamanifarizhandi2024oja} take the observation of maximal depth as a sample median and establish its consistency.

Inference for depth has targeted the depth of a given point. Limit theorems for the empirical depth function and the depth process \citep{dumbgen1992limit,masse2004asymptotics} and bootstrap limit theory for U-statistics \citep{arcones1992bootstrap} support pointwise inference, and confidence intervals have been constructed from such limits and by subsampling \citep{ramsay2019integrated,jentsch2026resampling}. Depth is also used to build confidence regions for multivariate parameters \citep{yehsingh1997balanced,liusingh1997limiting} and simultaneous central regions for functional data \citep{narisetty2016extremal}. Uniform bounds on the depth estimation error \citep{clemencon2023affine} hold at all points at once, and hence at the observed objects, but they are conservative. None of this work says which observed objects could be the most central, which are among the least central, or where each one ranks, with a guarantee that holds jointly over the sample. We address this problem with the reference depth as the measure of centrality.

Confidence sets for ranks and for the best among a given set of candidates have a long history \citep{gupta1965,hsu1981,hansen2011mcs,mogstad2024inference} and are also used for model selection \citep{lei2020cvc,kim2025argmin}. In these works each candidate, such as a neighbourhood, a country or a statistical model, has a fixed parameter that the data estimate. Here the observations are both the candidates and the reference points and opponents used to compare them, so the guarantees of these methods do not apply. Like multiple comparisons with the best \citep{hsu1981} and the rank projections of \citet{mogstad2024inference}, our centre set tests each candidate through its differences in estimated depth from the other observations.

Ranking inference in the Bradley--Terry--Luce model \citep{bradley1952rank,gao2023uncertainty} assumes independent comparisons and a correctly specified model. In our setting the comparisons are induced by the data themselves, they share reference points and opponents, and no model is assumed. \citet{lyu2026core} rank objects by a Bradley--Terry--Luce projection of estimated pairwise preferences; when all pairs receive equal weight, its fitted scores increase with the estimated reference depths, so the present paper provides inference for this ranking.

Depth is also a standard tool for flagging outliers, by cutting the ordering at a fixed fraction of the sample or at a cut-off calibrated by the bootstrap \citep{dang2010nonparametric,febrero2008outliers}. Classical outlier identification rules control the probability of flagging a regular observation under a parametric model \citep{davies1993identification}, and conformal methods, including ones that use depth as the score, give error control when new points are compared with a clean reference sample \citep{bates2023testing,kimpark2026conformal}. Our outlier sets give error control within a single unlabelled sample.

\section{Setting}\label{sec:setting}

\subsection{The reference depth}\label{sec:mr}

Let $F$ be a probability measure on a space $\X$ equipped with a measurable dissimilarity $\delta:\X\times\X\to[0,\infty)$, and let $X_1,\dots,X_n$ be an i.i.d.\ sample from $F$. Given two objects, we ask which one tends to be closer to a typical draw from $F$. This leads to the population preference
\[
p_F(x,y)=\P_{Z\sim F}\{\delta(Z,x)>\delta(Z,y)\},
\]
the probability that a random reference point $Z$ finds $y$ closer than $x$, and to the reference depth
\begin{equation}\label{eq:r-def}
\rF(y)=\E_{X\sim F}\{p_F(X,y)\}=\P\{\delta(Z,X)>\delta(Z,y)\},
\end{equation}
where the opponent $X$ and the reference point $Z$ are independent draws from $F$. The function $1-\rF$ was introduced by \citet{liu2022quantiles}, whose global quantiles are defined through its ordering; in particular, the maximiser of $\rF$ is their metric median. Throughout we work in the no-ties regime, $\P\{\delta(Z,X)=\delta(Z,y)\}=0$ for every $y$ in the support of $F$, so that $p_F(x,y)+p_F(y,x)=1$ for $F\otimes F$-almost every pair $(x,y)$; we return to ties in Section~\ref{sec:discussion}. When $\delta$ is a metric, $\rF$ is unchanged by isometries, tends to zero at infinity and is continuous on the support of $F$, like the metric depths of \citet{geenens2023statistical}; its maximisers need not be a centre of symmetry, and for an equal mixture of two well-separated normal distributions they lie near the two modes (Supplementary Section~S1).

The reference depth at $X_i$ is estimated by the average win rate
\begin{equation}\label{eq:rhat}
\rh_i=\frac{1}{(n-1)(n-2)}\sum_{l\neq i}\ \sum_{j\notin\{i,l\}}\I\{\delta(X_l,X_j)>\delta(X_l,X_i)\}.
\end{equation}
Write $\omega_y(u,v)=\I\{\delta(v,u)>\delta(v,y)\}$ for the indicator that the reference point $v$ is closer to $y$ than to the opponent $u$, so that the summand in \eqref{eq:rhat} is $\omega_{X_i}(X_j,X_l)$. Conditionally on $X_i$, $\rh_i$ is an unbiased U-statistic of order two in the other $n-1$ observations,
\[
\rh_i=\binom{n-1}{2}^{-1}\sum_{\substack{j<l\\ j,l\neq i}}h_{X_i}(X_j,X_l),\qquad h_y(u,v)=\tfrac12\{\omega_y(u,v)+\omega_y(v,u)\},
\]
with $\E\,h_y(X_1,X_2)=\rF(y)$. The kernel $h_y$ averages the two comparisons in which $u$ and $v$ exchange roles. Other depths that are U-statistics of this kind, such as metric lens depth and metric spatial depth, are treated in Supplementary Section~S1.

\subsection{Inferential targets}\label{sec:targets}

Because the decisions concern observed objects, our targets are indexed by the sample. The most central observation $I^\star$ is the smallest index in $\argmax_{i\le n}\rF(X_i)$. The in-sample rank of $X_i$ is $R_i=1+\#\{j:\rF(X_j)>\rF(X_i)\}$, and the bottom-$k$ set is $\mathcal B_k=\{i:R_i>n-k\}$, which may have fewer than $k$ members if population depths tie. We seek a centre set $\Chat$, outlier sets $\widehat O_k$ and integers $R_i^{\rm lo}\le R_i^{\rm up}$ computed from the data such that, asymptotically, $\P(I^\star\in\Chat)\ge1-\alpha$, $\P(\widehat O_k\subseteq \mathcal B_k\ \text{for every }k)\ge1-\alpha$ and
\[
\P\big(R_i^{\rm lo}\le R_i\le R_i^{\rm up}\ \text{for all }i\le n\big)\ge1-\alpha.
\]

Since $\Chat$ contains $I^\star$ with asymptotic probability at least $1-\alpha$, any property that holds for every member of $\Chat$ also holds for the most central observation with at least this probability. The guarantee for $\widehat O_k$ concerns each of its members, all of which lie in the bottom $k$.

One could instead target the population centre $\mu=\argmax_y\rF(y)$, but a confidence set for $\mu$ need not name an observed object. For a distribution on $\R^d$ with a positive continuous density near $\mu$, the observation closest to $\mu$ is at distance of order $n^{-1/d}$, whereas in a regular location family a confidence ball for $\mu$ has radius of order $n^{-1/2}$. For $d\ge3$ the spacing between the observed objects near $\mu$ is therefore larger than the uncertainty about $\mu$, and a confidence ball for $\mu$ typically contains no observation (Section~\ref{sec:info}).

\section{Methodology}\label{sec:method}

Our task is to turn the estimated depths $\rh_1,\dots,\rh_n$ into the statements of Section~\ref{sec:targets}, which concern the unknown values $\rF(X_1),\dots,\rF(X_n)$ jointly. A simultaneous confidence band for these values yields all three statements at once.

\subsection{Simultaneous confidence bands}\label{sec:lemma}

A band is a family of intervals $[\rh_i-w_i,\rh_i+w_i]$, $i\le n$, with data-dependent $w_i\ge0$. When all intervals contain their depths and the interval of $X_j$ lies entirely above that of $X_i$, then $X_j$ is deeper than $X_i$. Counting such comparisons bounds the rank of every object.

\begin{lemma}[From bands to rank statements]\label{lem:sets}
Let $w_1,\dots,w_n\ge0$ and define
\[
R_i^{\rm lo}=1+\#\{j:\rh_j-w_j>\rh_i+w_i\},\qquad R_i^{\rm up}=n-\#\{j:\rh_j+w_j<\rh_i-w_i\},
\]
$\Chat_{\rm band}=\{i:R_i^{\rm lo}=1\}$ and $\widehat O_k^{\rm band}=\{i:R_i^{\rm lo}>n-k\}$. On the event $\{|\rh_i-\rF(X_i)|\le w_i\ \text{for all }i\}$ we have $R_i^{\rm lo}\le R_i\le R_i^{\rm up}$ for all $i$, $I^\star\in\Chat_{\rm band}$ and $\widehat O_k^{\rm band}\subseteq \mathcal B_k$.
\end{lemma}

The lemma holds for any choice of the $w_i$. It separates the probabilistic question, whether the band covers all depths at once, from the combinatorial step of turning the band into rank statements. To choose the $w_i$ we need the joint distribution of the errors $\rh_i-\rF(X_i)$.

\subsection{Errors of the estimated depths}\label{sec:lin}

Each $\rh_i$ is a U-statistic of the other observations, so its error is approximately linear in them. Conditionally on $X_i=y$, the Hoeffding decomposition of \eqref{eq:rhat} gives
\begin{equation}\label{eq:hajek}
\rh_i-\rF(X_i)=\frac{1}{n-1}\sum_{j\neq i}\psi_{X_i}(X_j)+\rho_i,\qquad
\psi_y(u)=a_y(u)+b_y(u)-2\rF(y),
\end{equation}
where $\rho_i$ is a remainder and, with $X\sim F$, $a_y(u)=\E\,\omega_y(X,u)$ is the contribution of $u$ as a reference point and $b_y(u)=\E\,\omega_y(u,X)$ is its contribution as an opponent, so that $\E\,h_y(u,X)=\{a_y(u)+b_y(u)\}/2$. The remainder is uniformly small, $\max_i|\rho_i|=O_{\P}(\log n/n)$, by an exponential inequality for U-statistics \citep[Supplementary Section~S1;][]{houdre2003}. Here and below, $Q_n=O_{\P}(a_n)$ means that for every $\eta>0$ there is an $M$ with $\P(|Q_n|>Ma_n)\le\eta$ for all large $n$, and $o(1)$ denotes a deterministic term that tends to zero as $n\to\infty$. The leading term is an average of $n-1$ bounded variables, and its variance is $v(X_i)/(n-1)$ with $v(y)=\Var\{\psi_y(X)\}$.

We estimate $\psi_{X_i}$ by plugging in the empirical distribution of the other observations. For $j\neq i$ put
\begin{align*}
\hat a_i(j)&=\frac{1}{n-2}\sum_{l\notin\{i,j\}}\I\{\delta(X_j,X_l)>\delta(X_j,X_i)\},\\
\hat b_i(j)&=\frac{1}{n-2}\sum_{l\notin\{i,j\}}\I\{\delta(X_l,X_j)>\delta(X_l,X_i)\},
\end{align*}
and $\hat\psi_i(j)=\hat a_i(j)+\hat b_i(j)-2\rh_i$, with $\hat\psi_i(i)=0$. Both $\hat a_i(\cdot)$ and $\hat b_i(\cdot)$ average to $\rh_i$ over $j$, so the empirical influence function $\hat\psi_i$ of $\rh_i$ is centred. The same observation $X_j$ influences every estimated depth, through $\hat\psi_1(j),\dots,\hat\psi_n(j)$; this shared influence makes the estimated depths dependent, and the resampling schemes below reproduce the dependence by using the same draws for all objects.

The variance $v(y)$ can differ greatly across the sample. Since $a_y$ and $b_y$ take values in $[0,1]$ and both have mean $\rF(y)$, we have $v(y)\le4\rF(y)$, so $v(y)\to0$ along any sequence of points whose depth tends to zero. For $y$ near $\mu$, $v(y)$ is also small, because a point that favours $y$ as a reference point is a strong opponent of $y$, so $a_y(u)$ and $b_y(u)$ offset each other. At moderate distance from $\mu$ neither low depth nor this cancellation keeps $v(y)$ small.

\subsection{Multiplier and guarded bootstrap bands}\label{sec:band}

Since the precision of $\rh_i$ varies across objects, we take $w_i$ proportional to an estimated standard error $\hat s_i$ of $\rh_i$. Let $\hat v_i=(n-1)^{-1}\sum_j\hat\psi_i(j)^2$, fix a small floor $\underline v>0$ in advance and put $\hat s_i=\{\max(\hat v_i,\underline v)/(n-1)\}^{1/2}$. The floor stabilises studentisation for observations whose estimated influence variance is essentially zero; we use $\underline v=10^{-5}$.

The simplest such band, which we call the multiplier band, takes $w_i=\hat c\,\hat s_i$. Let $G_i=\sum_j\xi_j\hat\psi_i(j)$, where $\xi_1,\dots,\xi_n$ are independent $N(0,1)$ multipliers, drawn independently of the data and shared by all $i$. In all our procedures, the $p$-quantile of $B$ Monte Carlo values is the $\lceil(B+1)p\rceil$-th smallest of them. The critical value $\hat c$ is the $(1-\alpha)$-quantile of the values of
\begin{equation}\label{eq:Tstar}
T_n^\ast=\max_{i\le n}\frac{|G_i|}{(n-1)\hat s_i}
\end{equation}
over $B_{\rm m}$ independent draws of the multipliers. Its coverage tends to $1-\alpha$ (Theorem~\ref{thm:full}) but is lower in small samples (Section~\ref{sec:simband}), because the Gaussian approximation behind it is poor for objects whose depth is estimated from few wins. We therefore propose the guarded bootstrap band. It takes $w_i=c_{\rm g}\hat s_i$ with a critical value $c_{\rm g}$ from the bootstrap, which recomputes the estimated depths on resampled data, and it adds separate intervals for objects with few wins.

In each of $B_{\rm r}$ resamples, we draw $Y_1,\ldots,Y_n$ independently from the empirical distribution of the sample and use them as reference points and opponents for the original observations $X_1,\dots,X_n$. Define
\begin{align*}
\rh_i^*&=\binom n2^{-1}\sum_{j<l}h_{X_i}(Y_j,Y_l),&
\theta_{n,i}&=n^{-2}\sum_{j,l=1}^n h_{X_i}(X_j,X_l),\\
\hat\psi_i^*(j)&=\frac{2}{n-1}\sum_{l\ne j}h_{X_i}(Y_j,Y_l)-2\rh_i^*,&
(s_i^*)^2&=\frac{\max\{n^{-1}\sum_j\hat\psi_i^*(j)^2,\underline v\}}{n}.
\end{align*}
Repeated draws of the same observation count as distinct reference points and opponents. The quantity $\theta_{n,i}$ is the conditional mean of $\rh_i^*$, and $\hat\psi_i^*$ and $s_i^*$ are the bootstrap counterparts of $\hat\psi_i$ and $\hat s_i$.

The bootstrap approximates the joint distribution of the errors $\rh_i-\rF(X_i)$ by that of $\rh_i^*-\theta_{n,i}$. Conditionally on $X_i$, $\rh_i$ averages over the $(n-1)(n-2)/2$ pairs of the other observations. These pairs share observations, so their outcomes are dependent, but $\rh_i$ has tails no heavier than a proportion of wins in $L_n=\lfloor(n-1)/2\rfloor$ independent trials \citep[Supplementary Section~S1;][]{hoeffding1963}, so $L_n\rh_i$ acts as an effective number of wins. For an object of small depth this number is small and its distribution is skewed and discrete, and an object without wins has zero estimated influence variance.

Intervals from these binomial tail bounds hold for every $n$, but they ignore the variance of each estimated depth. We therefore use them only for objects with at most a few effective wins, and call $X_i$ shallow if $\rh_i\le\tau_n=\log\{2\log(en)\}/L_n$, that is, if it has at most $\log\{2\log(en)\}$ effective wins. We leave the shallow objects out of the bootstrap maximum, $M^*_{\rm g}=\max_{i:\rh_i>\tau_n}|\rh_i^*-\theta_{n,i}|/s_i^*$, with an empty maximum equal to zero, and let $c_{\rm g}$ be the $(1-\alpha/2)$-quantile of the $B_{\rm r}$ values of $M^*_{\rm g}$, capped at $K_n=\{2\log(4n/\alpha)\}^{1/2}$, an upper bound on the normal Bonferroni critical value for $n$ two-sided intervals at level $\alpha/2$. The cap binds only in small samples, where it limits the width of the band.

For the shallow objects we use the binomial tail bounds. Using the binary relative entropy $\mathrm{kl}(q,p)=q\log(q/p)+(1-q)\log\{(1-q)/(1-p)\}$ for $q,p\in[0,1]$, with the usual conventions at the endpoints, we put
\begin{equation}\label{eq:guard}
\begin{split}
[\ell_n(q),u_n(q)]&=\{p\in[0,1]:L_n\,\mathrm{kl}(q,p)\le\log(4n/\alpha)\},\\
w_i^{\rm g}&=\max\left\{c_{\rm g}\hat s_i,\ \I\{\rh_i\le\tau_n\}
\max\{\rh_i-\ell_n(\rh_i),u_n(\rh_i)-\rh_i\}\right\}.
\end{split}
\end{equation}
The interval $[\ell_n(\rh_i),u_n(\rh_i)]$ contains $\rF(X_i)$ with probability at least $1-\alpha/(2n)$ for every $n$, so these intervals hold jointly with probability at least $1-\alpha/2$. We call $\rh_i\pm w_i^{\rm g}$ the \emph{guarded bootstrap band}. It has asymptotic coverage at least $1-\alpha$ under the conditions of Section~\ref{sec:valid} (Theorem~\ref{thm:guard}). Lemma~\ref{lem:sets} with $w_i=w_i^{\rm g}$ gives its rank intervals, its centre set $\Chat_{\rm band}$ and its outlier sets $\widehat O_k^{\rm band}$.

\subsection{Centre sets}\label{sec:row}

The band centre set $\Chat_{\rm band}$ of Section~\ref{sec:band} has asymptotic coverage under the conditions of Section~\ref{sec:valid}, but it is large, because it compares objects only through intervals for their individual depths. Whether $X_i$ could be the most central observation is a question about the differences $\rF(X_j)-\rF(X_i)$. For nearby objects the estimated differences $\rh_j-\rh_i$ are much more precise than the individual estimated depths, because both depths are estimated with the same opponents and the same reference points, so much of their estimation error cancels in the difference. We therefore base the centre set on differences. Viewing $\hat\psi_i$ as the vector $(\hat\psi_i(1),\dots,\hat\psi_i(n))$, let $s_{ij}=\|\hat\psi_i-\hat\psi_j\|_2$, so that $s_{ij}/(n-1)$ is an estimated standard error of $\rh_j-\rh_i$. Since $\hat\psi_i(i)=\hat\psi_j(j)=0$, coordinates $i$ and $j$ of $\hat\psi_i-\hat\psi_j$ are $-\hat\psi_j(i)$ and $\hat\psi_i(j)$, the influence of each of the two objects on the estimated depth of the other; we call them the cross-influence coordinates. We keep them; they preserve the coverage of $\Chat$ on the line when $\mu$ is not a median of $F$ (Section~\ref{sec:simcentre}). With the multiplier sums $G_i$ of Section~\ref{sec:band}, $G_j-G_i$ is the multiplier counterpart of the estimation error of $(n-1)(\rh_j-\rh_i)$.

For each candidate $X_i$ we test the hypothesis $H_i$ that it is the most central observation, using the other observations as its competitors; $\Chat$ keeps the candidates that are not rejected. The observed and multiplier contrasts are
\[
\hat Z_{ij}=\frac{(n-1)(\rh_j-\rh_i)}{\max(s_{ij},\epsilon)},\qquad Z^\ast_{ij}=\frac{G_j-G_i}{\max(s_{ij},\epsilon)},\qquad j\neq i,
\]
where $\epsilon=10^{-12}$ only prevents division by zero. Under $H_i$ every population-depth difference $\rF(X_j)-\rF(X_i)$ is non-positive. Most competitors are clearly worse than a central candidate, and including them only inflates the critical value of the test. Following the generalised moment selection of \citet{andrewssoares2010}, we keep the competitors whose estimated depths could affect the test,
\[
\mathcal U_i=\{j\neq i:\hat Z_{ij}\ge-\kappa_n\},\qquad\kappa_n=\sqrt{\log n},
\]
and let $c_i$ be the $(1-\alpha)$-quantile of the values of $\max\{0,\max_{j\in \mathcal U_i}Z^\ast_{ij}\}$ over $B_{\rm m}$ independent draws of the multipliers. The centre set is
\begin{equation}\label{eq:Crow}
\Chat=\Big\{i:\ \max\Big\{0,\max_{j\in \mathcal U_i}\hat Z_{ij}\Big\}\le c_i\Big\}.
\end{equation}
Competitors outside $\mathcal U_i$ have negative contrasts and cannot cause a rejection, and a candidate with no competitor in $\mathcal U_i$ is never rejected.

The event $\{I^\star\in\Chat\}$ involves only the test for the single index $I^\star$, so the procedure needs no correction for multiplicity across candidates. Sections~\ref{sec:1d} and \ref{sec:info} establish the validity of this test at the random index $I^\star$ for Euclidean data and, under the conditions of Remark~\ref{rem:metric}, for metric spaces of high or low intrinsic dimension.

\subsection{Outlier sets}\label{sec:outlier}

The band outlier sets $\widehat O_k^{\rm band}$ of Section~\ref{sec:band} also compare intervals for single depths, so we again use differences. Unlike the centre set, which need only contain one particular object, an outlier set must contain only objects in the bottom $k$. Since $X_i$ lies in $\mathcal B_k$ once $n-k$ other objects are deeper, we flag $X_i$ when at least $n-k$ other objects are significantly deeper. For these comparisons we use the variance floor $\underline v$ of Section~\ref{sec:band} and put
\[
s_{ij}^{(\underline v)}=\max\{s_{ij},\sqrt{(n-1)\underline v}\},\qquad
 t_{ij}^{(\underline v)}=\frac{(n-1)(\rh_j-\rh_i)}{s_{ij}^{(\underline v)}}.
\]
The outlier sets compare all ordered pairs at once, including pairs of nearly identical objects, for which both $\rh_j-\rh_i$ and $s_{ij}$ are close to zero; the floor $\underline v$ keeps such pairs from being declared significantly different. The centre set uses no such floor, because its precision comes from exactly these close pairs. Let $c_{\rm all}$ be the $(1-\alpha)$-quantile of the $B_{\rm m}$ values of $\max_{i\neq j}(G_j-G_i)/s_{ij}$, the maximum over all ordered pairs. A pair with $s_{ij}=0$ contributes zero to this maximum. The outlier set is
\begin{equation}\label{eq:Opair}
\widehat O_k=\big\{i:\ \#\{j:t_{ij}^{(\underline v)}>c_{\rm all}\}\ge n-k\big\}.
\end{equation}

On the event that $(\rh_j-\rh_i)-\{\rF(X_j)-\rF(X_i)\}\le c_{\rm all}s_{ij}^{(\underline v)}/(n-1)$ for every ordered pair, each flagged object has at least $n-k$ deeper objects and therefore lies in $\mathcal B_k$, for all $k$ at once. Since $c_{\rm all}$ is computed from $s_{ij}$, which is at most $s_{ij}^{(\underline v)}$, the floor only widens the bounds in this event. Theorem~\ref{thm:outlier} shows that $\widehat O_k\subseteq \mathcal B_k$ for every $k$ with asymptotic probability at least $1-\alpha$. Supplementary Section~S7 summarises the computation of the rank intervals, the centre set and the outlier sets in an algorithm and gives its cost.

\section{Theory}\label{sec:theory}

The procedures of Section~\ref{sec:method} raise two questions. The first is whether they are valid, given that the targets $\rF(X_i)$ are indexed by the same sample that is used to estimate them. The second is how informative the centre set is: how many objects it contains, and whether another procedure could do better. Section~\ref{sec:valid} answers the first question for the bands, for the band centre sets $\Chat_{\rm band}$ and band outlier sets $\widehat O_k^{\rm band}$ that Lemma~\ref{lem:sets} derives from them, and for the outlier sets $\widehat O_k$. The centre set $\Chat$ is not derived from a band, and Sections~\ref{sec:1d} and \ref{sec:info} establish its coverage together with its size. Supplementary Table~S1 summarises the answers.

\subsection{Validity of the bands and outlier sets}\label{sec:valid}

The critical value of a simultaneous band comes from a maximum over the $n$ observed objects, so the objects indexing the terms of this maximum are random. To control this maximum we need a central limit theorem for the influence functions $\psi_y$ that holds uniformly over all possible objects $y$. The influence functions are built from the comparison indicators $\omega_y$ of Section~\ref{sec:mr} through $a_y$ and $b_y$. The uniform central limit theorem holds when these indicators, as functions of the pair $(u,v)$, form a Vapnik--Chervonenkis (VC) class, that is, a class of limited combinatorial complexity \citep[Section~2.6]{vdvw1996}. With the floor $\underline v$ of Section~\ref{sec:band}, the standardised influence functions and their class are
\[
\tilde\psi_y=\frac{\psi_y}{\{\max(v(y),\underline v)\}^{1/2}},\qquad \mathcal F=\{\tilde\psi_y:\ y\in\mathrm{supp}\,F\}.
\]

\begin{assumption}\label{ass:full}
\begin{enumerate}[label=(\alph*),ref=\alph*,leftmargin=*]
\item\label{ass:full:space} $(\X,\delta)$ is a Polish metric space, $X_1,\dots,X_n$ are i.i.d.\ from $F$, and $\P\{\delta(Z,X)=\delta(Z,y)\}=0$ for every $y\in\mathrm{supp}\,F$.
\item\label{ass:full:entropy} The comparison indicators $\omega_y$, $y\in\X$, as functions of the pair $(u,v)$, form a VC class, and $(y,u,v)\mapsto \omega_y(u,v)$ is jointly Borel measurable.
\item\label{ass:full:nondegeneracy} $v(y)>0$ for some $y\in\mathrm{supp}\,F$.
\end{enumerate}
\end{assumption}

Condition~(\ref{ass:full:space}) excludes ties in the comparisons, and condition~(\ref{ass:full:nondegeneracy}) excludes the trivial case in which every influence function vanishes. Condition~(\ref{ass:full:entropy}) is the main restriction. It holds whenever $\delta(x,y)=\|\phi(x)-\phi(y)\|_p$ for a Borel map $\phi$ into $\R^d$ and some $p\ge1$, which covers Euclidean data and weighted $\ell_p$ distances between finite-dimensional features (Supplementary Section~S1). For other spaces, such as trees under a path-length metric, the VC requirement in condition~(\ref{ass:full:entropy}) can be replaced by the requirement that $\mathcal F$ and $\{\psi_y\}$ are $F$-Donsker and $\{\psi_y^2\}$ is $F$-Glivenko--Cantelli (Supplementary Section~S1).

Let $T_n=\max_{i\le n}|\rh_i-\rF(X_i)|/\hat s_i$ be the maximum standardised error over the observed objects, and let $\mathbb G$ be a centred Gaussian process on $\mathcal F$ with covariance $\E\{\tilde\psi_y(X)\tilde\psi_{y'}(X)\}$, tight in $\ell^\infty(\mathcal F)$, so that $\mathbb G(\tilde\psi_y)$ is the limit of the standardised error $\{\rh_i-\rF(X_i)\}/\hat s_i$ of an observation at $X_i=y$.

\begin{theorem}[Validity of the multiplier band]\label{thm:full}
Let Assumption~\ref{ass:full} hold, fix $0<\alpha<1$ and $\underline v>0$, and let $B_{\rm m}\to\infty$. Then
\begin{enumerate}[label=(\roman*),leftmargin=*]
\item $T_n\rightsquigarrow\|\mathbb G\|_{\mathcal F}=\sup_{y\in\mathrm{supp}\,F}|\mathbb G(\tilde\psi_y)|$, where $\rightsquigarrow$ denotes convergence in distribution;
\item the distribution of $\|\mathbb G\|_{\mathcal F}$ is continuous, and $\sup_t|\P_\xi(T_n^\ast\le t)-\P(\|\mathbb G\|_{\mathcal F}\le t)|\to0$ in probability, where $\P_\xi$ is the probability over the multipliers given the data;
\item $\P\{|\rh_i-\rF(X_i)|\le\hat c\,\hat s_i\ \text{for all }i\le n\}\to1-\alpha$.
\end{enumerate}
Consequently, with $w_i=\hat c\,\hat s_i$ in Lemma~\ref{lem:sets}, the statements $R_i^{\rm lo}\le R_i\le R_i^{\rm up}$ for all $i$, $I^\star\in\Chat_{\rm band}$ and $\widehat O_k^{\rm band}\subseteq \mathcal B_k$ for every $k$ hold jointly with probability at least $1-\alpha+o(1)$.
\end{theorem}

Theorem~\ref{thm:full} shows that the random indexing has no effect in the limit: the sample fills the support, so the maximum over the observed objects behaves like the supremum over all objects (Supplementary Section~S1). The critical value therefore converges to a constant, the $(1-\alpha)$-quantile of $\|\mathbb G\|_{\mathcal F}$, instead of growing like $(2\log n)^{1/2}$ as it would for $n$ independent statistics.

\begin{theorem}[Validity of the guarded bootstrap band]\label{thm:guard}
Under Assumption~\ref{ass:full}, with fixed $0<\alpha<1$ and $\underline v>0$ and with $B_{\rm r}\to\infty$, the band in \eqref{eq:guard} has asymptotic simultaneous coverage at least $1-\alpha$, and its coverage tends to $1-\alpha/2$. In particular, with $w_i=w_i^{\rm g}$ in Lemma~\ref{lem:sets}, the statements $R_i^{\rm lo}\le R_i\le R_i^{\rm up}$ for all $i$, $I^\star\in\Chat_{\rm band}$ and $\widehat O_k^{\rm band}\subseteq \mathcal B_k$ for every $k$ hold jointly with probability at least $1-\alpha+o(1)$.
For every $n\ge3$ and any i.i.d.\ sample,
\[
\P\{|\rh_i-\rF(X_i)|\le w_i^{\rm g}\ \text{for all }i\text{ with }\rh_i\le\tau_n\}\ge1-\alpha/2.
\]
\end{theorem}
The intervals for shallow objects matter only in finite samples. Leaving the shallow objects out of the bootstrap maximum $M^*_{\rm g}$ does not change its limit, so $c_{\rm g}$ tends to the $(1-\alpha/2)$-quantile of $\|\mathbb G\|_{\mathcal F}$, and for large $n$ the intervals for shallow objects are narrower than $c_{\rm g}\hat s_i$, so that $w_i^{\rm g}=c_{\rm g}\hat s_i$ for every $i$; by Theorem~\ref{thm:full} the coverage therefore tends to $1-\alpha/2$ (Supplementary Section~S1).

The outlier set $\widehat O_k$ compares differences of estimated depths, and its proof applies the band argument to differences of influence functions (Supplementary Section~S1). With the fixed variance floor $\underline v$, the critical value $c_{\rm all}$ controls every outlier set at once.
\begin{theorem}[Outlier sets with a variance floor]\label{thm:outlier}
Let Assumption~\ref{ass:full} hold, fix $0<\alpha<1$ and $\underline v>0$, and let $B_{\rm m}\to\infty$. Then
\[
\liminf_{n\to\infty}\P\{\widehat O_k\subseteq \mathcal B_k
\ \text{for every }k=1,\ldots,n\}\ge1-\alpha.
\]
\end{theorem}

\subsection{Centre sets in one dimension}\label{sec:1d}

We study the centre set for the reference depth in Euclidean space through the test for the most central observation, and we determine at the same time how many objects it contains. We start with the real line, where the influence function has a simple explicit form and where a lower bound over all procedures shows how good a given rate is. In this and the next subsection, $\Chat$ is computed as in Section~\ref{sec:row} with $\alpha\in(0,1/2)$ fixed and $B_{\rm m}\to\infty$.

Take $\delta(x,y)=|x-y|$ and let $F$ have a density $f$. Put $J=1-2F(\mu)$, which is zero when $\mu$ is a median of $F$. The one-dimensional results use the following conditions, stated precisely in Supplementary Section~S2.
\begin{assumption}\label{ass:1d}
\begin{enumerate}[label=(\alph*),ref=\alph*,leftmargin=*]
\item\label{ass:1d:density} $f$ is three times continuously differentiable on $\R$, with bounded and integrable derivatives, and bounded away from zero near $\mu$.
\item\label{ass:1d:maximum} $\sup_{|y-\mu|\ge\eta}\rF(y)<\rF(\mu)$ for every $\eta>0$, and $\rF''(\mu)<0$.
\end{enumerate}
\end{assumption}

Assumption~\ref{ass:1d} implies Assumption~\ref{ass:full} (Supplementary Section~S2). Whether $\Chat$ contains $I^\star$, and how many candidates it keeps, depend on how fast the influence functions $\psi_y$ change as $y$ moves near $\mu$. The reference-point part $a_y(u)$ is continuous in $y$. The opponent part $b_y(u)$ is the probability that a random reference point falls on the same side of the midpoint $(y+u)/2$ as $y$, so it jumps from $F(u)$ to $1-F(u)$ when $y$ crosses $u$. Near $\mu$ the jump is close to $J$, and Supplementary Section~S2 shows that, under Assumption~\ref{ass:1d},
\[
\|\psi_{\mu+t}-\psi_\mu\|_{L^2(F)}^2=f(\mu)J^2|t|+O(t^2),\qquad t\to0.
\]
When $J=0$, for instance when $F$ is symmetric, the increments have squared $L^2(F)$ norms of order $t^2$. Otherwise they behave like the increments of a Brownian motion near $\mu$.

\begin{theorem}[Centre set in one dimension]\label{thm:row1d}
Let Assumption~\ref{ass:1d} hold. If $J=0$, then $\P(I^\star\in\Chat)\to1$ and there are constants $0<C_1<C_2<\infty$ such that $\P\{C_1(n\log n)^{1/2}\le|\Chat|\le C_2(n\log n)^{1/2}\}\to1$. If $J\neq0$, then $\P(I^\star\in\Chat)\to1-\alpha$ and $|\Chat|=O_{\P}\{n^{2/3}(\log n)^{1/3}\}$.
\end{theorem}

When $J=0$ the largest observed contrast of $I^\star$, $\max_{j\in\mathcal U_{I^\star}}\hat Z_{I^\star j}$, is $O_{\P}(\log\log n)$, whereas its critical value $c_{I^\star}$ grows like $(\log n)^{1/2}$, so the coverage tends to one. When $J\neq0$ both grow like $(2\log\log n)^{1/2}$ and have the same limit after normalisation, so the coverage tends to $1-\alpha$ (Supplementary Section~S4).

The next result bounds from below the size of any set whose coverage holds uniformly over a neighbourhood of the true location, in a location family where the shape of $F$ is known.

\begin{theorem}[Lower bound]\label{thm:1dlower}
For $\theta\in\R$ let $F_\theta$ be the distribution of $X+\theta$ with $X\sim F$, where $F$ satisfies Assumption~\ref{ass:1d} and has finite Fisher information for location, and let $I^\star_\theta$ be the most central observation under $F_\theta$. Write $\P_\theta$ and $\E_\theta$ for probability and expectation when the sample is drawn from $F_\theta$. Let $A_n\subseteq\{1,\dots,n\}$ be the set produced by any, possibly randomised, procedure such that
\[
\inf_{|\theta|\le\zeta/\sqrt n}\P_\theta(I^\star_\theta\in A_n)\ge1-\alpha-o(1)
\]
for some $\zeta>0$. Then $\liminf_n\E_0|A_n|/\sqrt n>0$, and $\liminf_n\P_0(|A_n|\ge C\sqrt n)>0$ for some $C>0$. The order $\sqrt n$ is attained in this family by a procedure that knows the shape of $F$.
\end{theorem}

Theorem~\ref{thm:1dlower} holds because, when $|\theta|$ is of order $n^{-1/2}$, no test based on the sample can reliably distinguish $F_\theta$ from $F$, whereas $I^\star_\theta$ moves across order $\sqrt n$ observations as $\theta$ ranges over these values. A set that contains $I^\star_\theta$ with probability at least $1-\alpha$ for all these $\theta$ must therefore contain a positive fraction of these observations. The centre set $\Chat$ is translation equivariant, so its coverage in Theorem~\ref{thm:row1d} is the same under every $F_\theta$, and Theorem~\ref{thm:1dlower} applies to it. When $J=0$, $|\Chat|$ is of order $(n\log n)^{1/2}$ by Theorem~\ref{thm:row1d}, so $\Chat$ is optimal up to a factor $(\log n)^{1/2}$. When $J\neq0$, the upper bound $n^{2/3}(\log n)^{1/3}$ in Theorem~\ref{thm:row1d} exceeds $\sqrt n$ by a factor $n^{1/6}(\log n)^{1/3}$, which comes from the jump in $b_y$. Supplementary Section~S2 gives the proof of Theorem~\ref{thm:1dlower}.

\subsection{Centre sets in higher dimensions and metric spaces}\label{sec:info}

The size of $\Chat$ is set by two distances near $\mu$: the distance within which candidates cannot be distinguished from $I^\star$, and the spacing of the observations. On the line this distance is, up to logarithmic factors, of order $n^{-1/2}$ when $J=0$ and at most of order $n^{-1/3}$ when $J\neq0$, because of the jump in $b_y$, while the spacing is of order $n^{-1}$, so $\Chat$ grows with $n$. In dimension $d\ge2$ a candidate $y$ generically does not cross an observation, so this jump does not arise, and the spacing near $\mu$ is of order $n^{-1/d}$. In higher dimensions we use the following conditions, stated precisely in Supplementary Section~S3.
\begin{assumption}\label{ass:highd}
$\X=\R^d$ with fixed $d\ge2$ and $\delta(x,y)=\|x-y\|$. The distribution $F$ has a bounded density $f$, and near $\mu$ the density $f$ is bounded away from zero and both $f$ and $\rF$ are three times continuously differentiable. The Hessian $\nabla^2\rF(\mu)$ is negative definite, and $\sup_{\|y-\mu\|\ge\eta}\rF(y)<\rF(\mu)$ for every $\eta>0$. The projection regularity conditions of Supplementary Section~S3 hold, which concern the half-space probabilities $\P\{\vartheta\trans(X-\mu)<s\}$ for unit vectors $\vartheta$ and thresholds $s$, the derivatives of $a_y(u)$ in $y$, and the probability that a comparison changes when the candidate moves.
\end{assumption}
Assumption~\ref{ass:highd} implies Assumption~\ref{ass:full} (Supplementary Section~S3). Under Assumption~\ref{ass:highd} the influence functions vary smoothly near $\mu$: for $y$ and $z$ near $\mu$, with $t=\|y-z\|$,
\[
\|\psi_y-\psi_z\|_{L^2(F)}\le\begin{cases}Ct\{\log(e/t)\}^{1/2},&d=2,\\ Ct,&d\ge3,\end{cases}
\]
so candidates that cannot be distinguished from $I^\star$ again lie within distance of order $n^{-1/2}$ of $\mu$, up to logarithmic factors. This distance is comparable to the spacing in the plane and smaller than it for $d\ge3$, so $\Chat$ contains at most a polylogarithmic number of observations in the plane and, with probability tending to one, only $I^\star$ for $d\ge3$ (Theorem~\ref{thm:highd}).

In the plane the opponent part $b_y(u)$ is the probability that a random reference point is closer to $y$ than to $u$. For $y$ and $u$ near $\mu$ it is close to $\P\{(u-y)\trans(X-\mu)<0\}$, the probability of a half-plane bounded by a line through $\mu$, and so depends on the direction of $u-y$. On the line there are only two directions, with probabilities $F(\mu)$ and $1-F(\mu)$, and their difference is the jump $J$. In the plane $b_y(u)$ likewise varies with the direction, unless every half-plane bounded by a line through $\mu$ has probability $1/2$. We call this the half-plane condition. It is the planar analogue of $J=0$, holds for distributions symmetric about $\mu$, and reduces the bound in Theorem~\ref{thm:highd}(i) from $(\log n)^2$ to $\log n$.

\begin{theorem}[Centre set in dimension $d\ge2$]\label{thm:highd}
Let Assumption~\ref{ass:highd} hold.
\begin{enumerate}[label=(\roman*),leftmargin=*]
\item If $d=2$, then $|\Chat|=O_{\P}\{(\log n)^2\}$, and $|\Chat|=O_{\P}(\log n)$ under the half-plane condition.
\item If $d\ge3$, then $\P(\Chat\neq\{I^\star\})\le C(\log n)^{1/2}n^{1/d-1/2}+o(1)$; in particular $\Chat=\{I^\star\}$ with probability tending to one.
\end{enumerate}
\end{theorem}

For $d\ge3$, Theorem~\ref{thm:highd}(ii) implies that $\Chat$ contains $I^\star$ with probability tending to one. In the plane $\Chat$ still contains several observations near $\mu$, so its coverage depends on the tests of $I^\star$ against these nearby competitors. Theorem~\ref{thm:plane} gives coverage under further conditions. When the half-plane condition holds, these include $\rF(\mu)\neq3/4$, which keeps the cross-influence coordinates of Section~\ref{sec:row} from vanishing for close pairs near $\mu$, so that $s_{ij}$ stays bounded away from zero for these pairs (Supplementary Section~S3).
\begin{theorem}[Coverage in the plane]\label{thm:plane}
Let $d=2$ and assume the conditions of Theorem~\ref{thm:highd}, together with the local entropy condition of Supplementary Section~S3 and, when the half-plane condition holds, $\rF(\mu)\neq3/4$ and the non-degeneracy condition of that section. Then $\liminf_n\P(I^\star\in\Chat)\ge1-\alpha$, with limit $1-\alpha$ when the half-plane condition fails.
\end{theorem}
The spherical normal distribution satisfies all the conditions of Theorems~\ref{thm:highd} and \ref{thm:plane} (Supplementary Section~S3), and Supplementary Section~S4 gives their proofs.

\begin{remark}[Metric spaces]\label{rem:metric}
Local probability mass also determines the size of the centre set beyond Euclidean space. For a distribution $F$ on a Polish metric space, suppose that $\sup_{\delta(y,\mu)\ge\eta}\rF(y)<\rF(\mu)$ for every $\eta>0$ and that, for some constants $0<\nu_1\le \nu_2<\infty$ and $D>0$,
\[
\nu_1\delta(y,\mu)^2\le\rF(\mu)-\rF(y)\le \nu_2\delta(y,\mu)^2,\qquad \nu_1t^D\le\P\{\delta(X,\mu)<t\}\le \nu_2t^D
\]
for $y$ near $\mu$ and small $t>0$. The exponent $D$ is the intrinsic dimension of $F$ at $\mu$. Assume also that $\|\psi_y-\psi_z\|_{L^2(F)}\le C\delta(y,z)^\gamma$ locally, where $C<\infty$ and $0<\gamma\le1$ are fixed. The exponent $\gamma$ describes how smoothly the influence functions vary with the candidate. Then $\Chat$ has size $O_{\P}\{1+n(\log n/n)^{D/[2(2-\gamma)]}\}$.

For singleton recovery, let $\Delta_n$ be the gap between the two largest of $\rF(X_1),\dots,\rF(X_n)$, and suppose that $\P\{\Delta_n\le \lambda_n n^{-2/D}\}\to0$ for every deterministic $\lambda_n\downarrow0$. Then $\Chat=\{I^\star\}$ with probability tending to one when $D>2(2-\gamma)$, and for $0<D<2$ the coverage of $\Chat$ still tends to one under further conditions on the local increments of the influence functions and kernels, which exclude $J\neq0$ on the line. Supplementary Section~S3 gives the precise conditions and proofs.
\end{remark}

\section{Simulation studies}\label{sec:sim}

All simulations use the same eight distributions: the $t_3$ and lognormal distributions on the line; gamma coordinates in $\R^2$, a $t_3$ distribution in $\R^5$, exponential coordinates in $\R^{50}$ and uniform coordinates in $\R^{1000}$, under the Euclidean distance; the von Mises--Fisher distribution on the sphere $S^{10}$ under the great-circle distance; and random distributions under the 1-Wasserstein distance. The sample sizes are $n=100$, 400 and 1600, with 1000 data sets for each distribution and sample size. For each distribution we drew one population of $100{,}000$ objects and computed the reference depth of every population object exactly by enumeration. Samples were drawn without replacement from the population, so each estimated depth is unbiased for the population depth of its object, and $I^\star$, the ranks $R_i$ and the bottom-$k$ sets are known. Throughout, $\alpha=0.05$, the multiplier procedures use $B_{\rm m}=999$ Gaussian draws and the bootstrap bands use $B_{\rm r}=300$ resamples. Supplementary Section~S5 gives the details and the results for every setting.

\subsection{Bands and outlier sets}\label{sec:simband}

\begin{figure}[t]
\centering
\includegraphics[width=\textwidth]{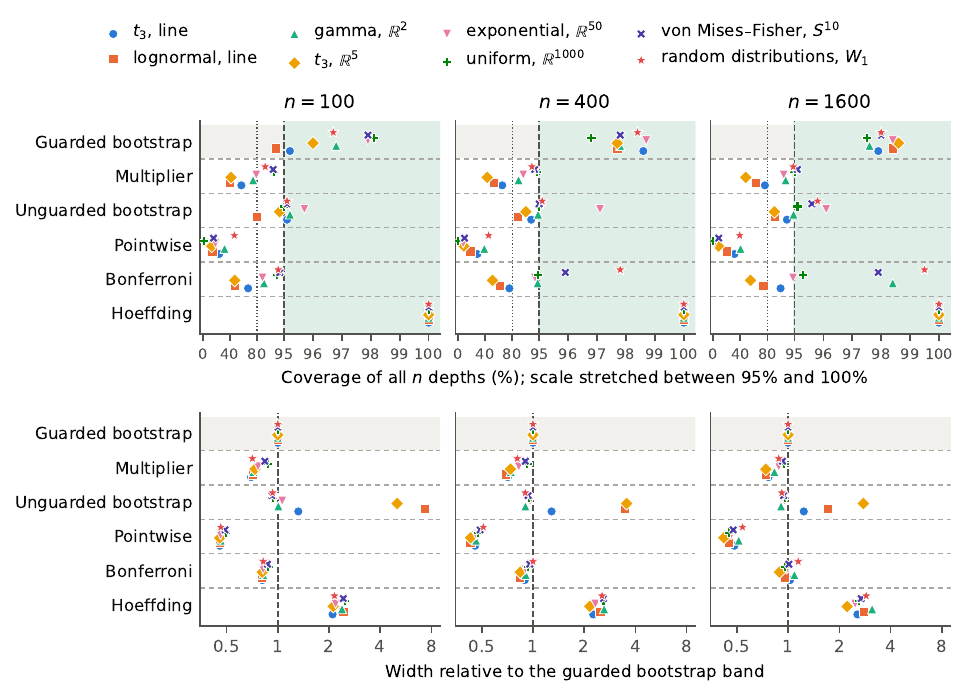}
\caption{Coverage of all $n$ depths (top; the scale is stretched between 95\% and 100\%, and the shaded region marks coverage of at least 95\%) and width relative to the guarded bootstrap band (bottom, log scale) for the six bands, with one point for each distribution; 1000 data sets per setting. Width is the median half-width across objects divided by that of the guarded bootstrap band in the same data set, with the median taken over data sets.}\label{fig:bands}
\end{figure}

We compared the guarded bootstrap band with five other bands: the multiplier band of Section~\ref{sec:band}; the unguarded bootstrap band, which drops the separate intervals for shallow objects and the cap and uses the $(1-\alpha)$-quantile of the bootstrap maximum over all objects; pointwise intervals $\rh_i\pm z_{1-\alpha/2}\hat s_i$, the form given by limit theory for the depth of a fixed point \citep{ramsay2019integrated}, with $z_p$ the standard normal $p$-quantile; Bonferroni intervals, which use $z_{1-\alpha/(2n)}$; and the Hoeffding band $\rh_i\pm\{\log(2n/\alpha)/(2L_n)\}^{1/2}$, which is valid for every $n$. Figure~\ref{fig:bands} shows the coverage and width of each band for every distribution and sample size, and Supplementary Section~S5 gives the values.

Of the six bands, only the guarded bootstrap band and the Hoeffding band attained at least 95\% coverage in every setting with $n\ge400$. The guarded bootstrap band covered all $n$ depths in at least 90.3\% of data sets at $n=100$, where the lowest value was for the lognormal distribution, and in at least 96.8\% for $n\ge400$, close to the limit of 97.5\% in Theorem~\ref{thm:guard}. The multiplier band was narrower but covered in only 40.3--95.1\% of data sets; its coverage increased with $n$ and was lowest for the two distributions on the line and the $t_3$ distribution in $\R^5$. The unguarded bootstrap band covered in 80.1--97.1\% of data sets, and its median half-width was up to 7.35 times that of the guarded bootstrap band. Pointwise intervals covered all depths in at most 46.5\% of data sets, and Bonferroni intervals in 47.1--99.5\%. The Hoeffding band covered all depths in every data set, but it was 2.1--3.1 times as wide as the guarded bootstrap band.

For the outlier sets we compared $\widehat O_k$ with the band outlier sets $\widehat O_k^{\rm band}$ of the multiplier band and the guarded bootstrap band, for $k=5$ and $k=n/20$. Table~\ref{tab:outlier} gives their power. No set flagged an object outside $\mathcal B_k$ in any data set, and $\widehat O_k$ had the highest power in every column.

\begin{table}[t]
\centering
\caption{Power of the outlier sets: percentage of the members of $\mathcal B_k$ flagged, averaged over data sets and over the eight distributions; 1000 data sets per setting.}\label{tab:outlier}
\small
\begin{tabular}{lccccc}
\toprule
 & $n=100$ & \multicolumn{2}{c}{$n=400$} & \multicolumn{2}{c}{$n=1600$}\\
\cmidrule(lr){2-2}\cmidrule(lr){3-4}\cmidrule(lr){5-6}
Outlier set & $k=5$ & $k=5$ & $k=20$ & $k=5$ & $k=80$\\
\midrule
$\widehat O_k$ & 13.6 & 20.1 & 45.7 & 31.0 & 67.8\\
$\widehat O_k^{\rm band}$, multiplier band & 6.3 & 9.9 & 22.3 & 17.0 & 47.0\\
$\widehat O_k^{\rm band}$, guarded bootstrap band & 0.1 & 1.7 & 10.5 & 8.6 & 39.9\\
\bottomrule
\end{tabular}
\end{table}

\subsection{Coverage and size of the centre sets}\label{sec:simcentre}

\begin{figure}[!t]
\centering
\includegraphics[width=0.95\textwidth]{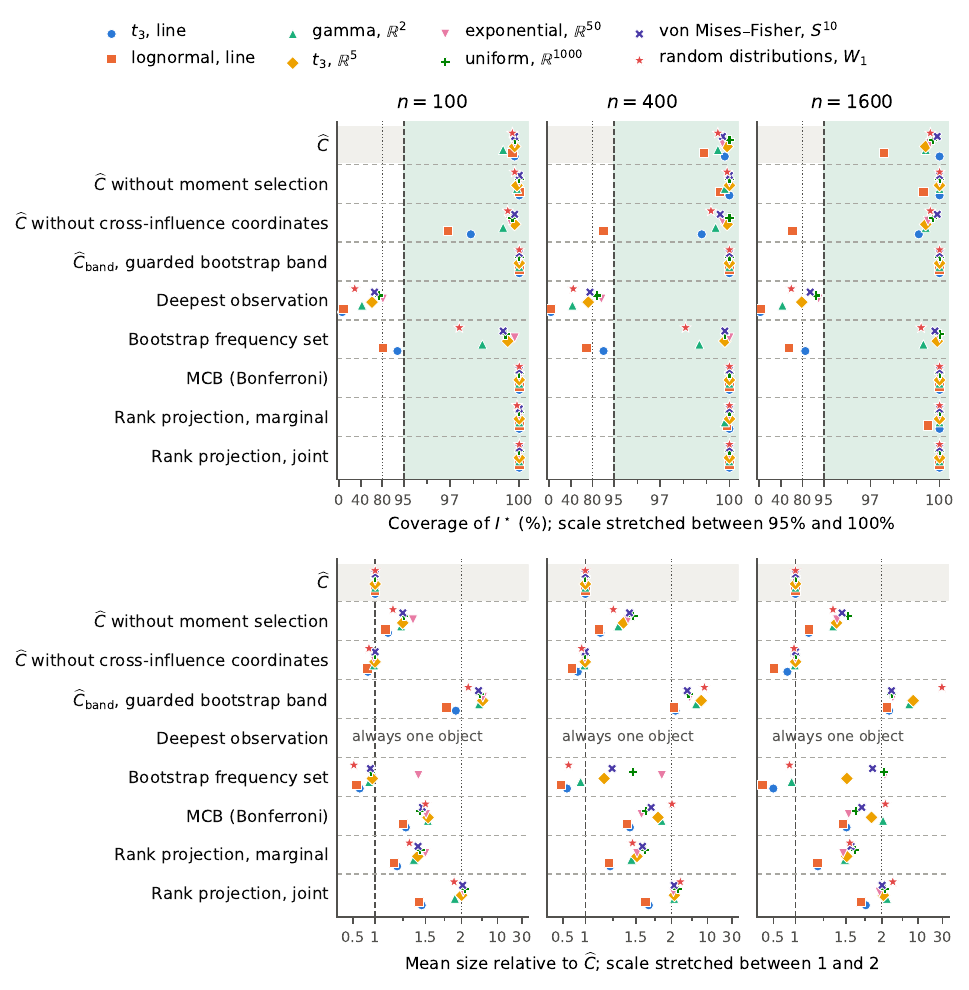}
\caption{Coverage of $I^\star$ (top; the scale is stretched between 95\% and 100\%, and the shaded region marks coverage of at least 95\%) and mean size relative to that of $\Chat$ (bottom; log scale, stretched between 1 and 2) for the nine centre sets, with one point for each distribution; 1000 data sets per setting. The deepest observation is a single object in every data set, so the bottom row shows no points for it.}\label{fig:centres}
\end{figure}

We compared $\Chat$ with eight other centre sets. Two variants each drop one of its components: $\Chat$ without moment selection takes $\mathcal U_i=\{j:j\neq i\}$ and so tests each candidate against all other observations, and $\Chat$ without cross-influence coordinates removes coordinates $i$ and $j$ from both $s_{ij}$ and $G_j-G_i$. The band centre set $\Chat_{\rm band}$ comes from the guarded bootstrap band. The five existing methods are the deepest observation $\argmax_i\rh_i$ alone; the bootstrap frequency set, which records the deepest observation in each of 499 bootstrap resamples and keeps the most frequent of these until they account for 95\% of the resamples; multiple comparisons with the best (MCB) of \citet{hsu1981} with a Bonferroni critical value, which keeps $X_i$ if $\max_{j\neq i}\hat Z_{ij}\le z_{1-\alpha/(n-1)}$; and the marginal and joint rank projections of \citet{mogstad2024inference}, which keep $X_i$ if no other observation is significantly deeper, that is, if $\max_{j\neq i}\hat Z_{ij}$ is at most the $(1-\alpha)$-quantile of $\max_{j\neq i}|Z^\ast_{ij}|$ (marginal) or of $\max_{j\neq l}|Z^\ast_{jl}|$ (joint). Figure~\ref{fig:centres} shows the coverage of $I^\star$ and the size of each set relative to $\Chat$ for every distribution and sample size, and Supplementary Section~S5 gives the details of the existing methods and the values.

The centre set $\Chat$ covered $I^\star$ in at least 97.6\% of data sets for every distribution and $n$ (Figure~\ref{fig:centres}, top). Its coverage was lowest for the lognormal distribution, for which $J\neq0$ and the limiting coverage is exactly 95\% (Theorem~\ref{thm:row1d}), and it fell from 99.7\% at $n=100$ to 97.6\% at $n=1600$.

Every other set that covered $I^\star$ in at least 95\% of data sets for every distribution and $n$ was larger than $\Chat$ (Figure~\ref{fig:centres}, bottom). Relative to $\Chat$, the mean size was 1.1--1.5 without moment selection, 1.2--3.3 for MCB and the two rank projections, and 1.8--29 for $\Chat_{\rm band}$. Without the cross-influence coordinates the set was 0.5--1.0 times as large as $\Chat$, but for the lognormal distribution, for which $J\neq0$, its coverage fell to 87.6\% at $n=400$ and 61.5\% at $n=1600$.

The deepest observation alone covered $I^\star$ in 1.8--8.0\% of data sets on the line and in 29.0--91.0\% for the other distributions, so the deepest observation alone is unreliable. The bootstrap frequency set covered $I^\star$ in at least 97.4\% of data sets for the distributions off the line, but on the line its coverage was 54.2--90.5\%, while its mean size was 0.35--0.61 times that of $\Chat$. The bootstrap frequency set estimates the sampling variability of the deepest observation, which differs from the uncertainty about $I^\star$.

\subsection{Growth of the centre sets}\label{sec:simgrowth}

\begin{figure}[t]
\centering
\includegraphics[width=\textwidth]{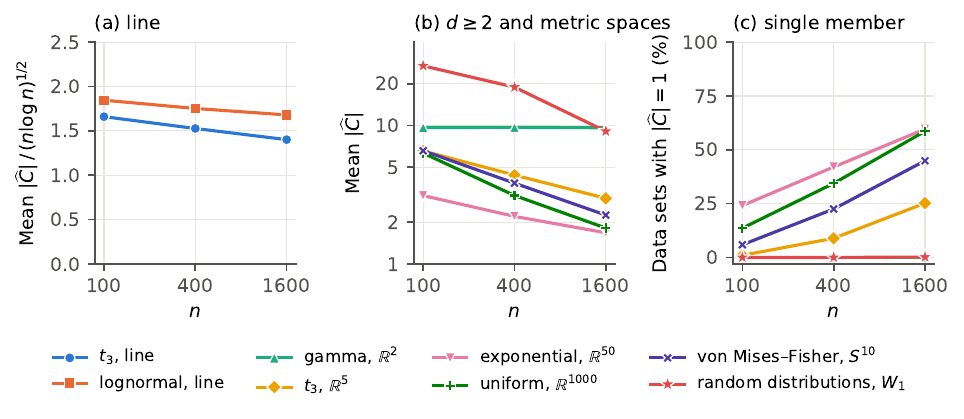}
\caption{(a) Mean size of $\Chat$ divided by $(n\log n)^{1/2}$ for the two distributions on the line; (b) mean size of $\Chat$ for the other six distributions, log--log scale; (c) percentage of data sets in which $\Chat$ has a single member, for the same six distributions; 1000 data sets per setting.}\label{fig:sizes}
\end{figure}

Figure~\ref{fig:sizes} compares the size of $\Chat$ with the rates of Theorems~\ref{thm:row1d} and \ref{thm:highd}. On the line the mean size of $\Chat$ was 1.4--1.8 times $(n\log n)^{1/2}$ for both distributions and every $n$ (Figure~\ref{fig:sizes}(a)). For the $t_3$ distribution, with $J=0$, this is the order given by Theorem~\ref{thm:row1d}. For the lognormal distribution, with $J\neq0$, the ratio decreased from 1.8 to 1.7 between $n=100$ and $n=1600$, so $|\Chat|$ grew more slowly than the upper bound $O_{\P}\{n^{2/3}(\log n)^{1/3}\}$ of Theorem~\ref{thm:row1d}, which would raise the ratio by a factor of about 1.5. For the gamma distribution in the plane the mean size stayed at about 10 for every $n$ (Figure~\ref{fig:sizes}(b)). For the three distributions in dimension $d\ge3$ and the distribution on the sphere the mean size decreased with $n$, and the percentage of data sets in which $\Chat$ had a single member rose with $n$, to 25--60\% at $n=1600$ (Figure~\ref{fig:sizes}(c)), in line with Theorem~\ref{thm:highd}(ii) and Remark~\ref{rem:metric}. For the random distributions the mean size fell from 27 to 9.

\section{Real data applications}\label{sec:data}

\subsection{Typical meteorological years}\label{sec:tmy}

For each station and calendar month, TMY3 chooses the candidate year in two steps \citep{wilcox2008}. First, it scores each candidate year by a weighted sum, over daily weather indices, of the mean absolute difference between the empirical distribution function of the index in that candidate year and in all candidate years together. The five candidate years with the smallest scores form the shortlist. Second, it selects one year from the shortlist after checking that its monthly means and medians are close to the long-term values and that it has no unusually long runs of extreme days. We asked which candidate years the data cannot distinguish from the most central one, and how much their heating or cooling degree days change among them. Degree days, the sums of daily temperature deviations from a base temperature, drive building energy use.

We used hourly data for 1973--2022 from the ISD-Lite files of the Integrated Surface Database (ISD, \url{https://www.ncei.noaa.gov/pub/data/noaa/isd-lite/}) of the US National Oceanic and Atmospheric Administration (NOAA). We analysed 40 stations drawn at random among those in the contiguous United States with records for the whole period, after setting aside the 65 stations examined while the method was developed. Each station and calendar month, or station-month, is analysed separately, with its candidate years as the sample. A candidate year is used if at least 90\% of its days have valid observations, and a station-month is analysed if it has at least 40 candidate years. This leaves 433 of the 480 station-months, so $n$ is between 40 and 50. Each candidate year is described by the empirical distribution functions of eight daily indices of dry-bulb temperature, dew point and wind speed, and the distance between two candidate years $i$ and $j$ of the same station-month is the TMY3 weighted sum computed between the two years,
\[
\delta(i,j)=\sum_{m=1}^8\varpi_m\,\frac1{|\mathcal G_m|}\sum_{x\in\mathcal G_m}|F_i^m(x)-F_j^m(x)|,
\]
where $F_i^m$ is the empirical distribution function of index $m$ in candidate year $i$, $\mathcal G_m$ consists of all daily values of index $m$ in the station-month, and the $\varpi_m$ are weights adapted from TMY3. Supplementary Section~S6 gives the details of the data processing.

In each station-month we computed $\Chat$ and, for comparison, MCB and the two rank projections, which kept their coverage in Section~\ref{sec:simcentre}. We compared these sets with the TMY3 shortlist and call its first member the first-ranked candidate year. The choice among candidate years matters for a building simulation when their degree days differ. With ${\rm DD}_i$ the heating (October to April) or cooling (May to September) degree days of candidate year $i$ and $\overline{\rm DD}$ their mean over the station-month, we measured the spread of degree days over a set $A$ of candidate years by $W(A)=(\max_{i\in A}{\rm DD}_i-\min_{i\in A}{\rm DD}_i)/\overline{\rm DD}$, and compared it with the median of $W(A)$ over 200 random sets of size $|\Chat|$.

\begin{table}[t]
\centering
\caption{Sets of candidate years in the 433 station-months. Each row is a set $A$ of candidate years, computed in every station-month. Size: median over station-months, with lower and upper quartiles in brackets. $|A|>5$ and Contains: percentage of station-months in which $A$ has more than five members, and in which it contains the first-ranked candidate year, the whole TMY3 shortlist and $\Chat$, respectively. $W(A)$: median over the 372 station-months with $\overline{\rm DD}\ge50$. The last three rows are references: the TMY3 shortlist, the deepest candidate year and random sets of the size of $\Chat$.}\label{tab:tmy}
\small
\setlength{\tabcolsep}{4pt}
\begin{tabular}{lcccccc}
\toprule
 & & & \multicolumn{3}{c}{Contains} & \\
\cmidrule(lr){4-6}
Set $A$ & Size & $|A|>5$ & first-ranked year & shortlist & $\Chat$ & $W(A)$\\
\midrule
$\Chat$ & 8 (7--10) & 85.2 & 99.5 & 51.5 & -- & 0.20\\
MCB (Bonferroni) & 12 (9--14) & 99.1 & 100 & 74.4 & 100 & 0.25\\
Rank projection, marginal & 11 (9--13) & 98.2 & 100 & 71.1 & 100 & 0.24\\
Rank projection, joint & 16 (13--18) & 100 & 100 & 91.0 & 100 & 0.30\\
\addlinespace
TMY3 shortlist & 5 & 0 & 100 & 100 & 3.9 & 0.14\\
Deepest candidate year & 1 & 0 & 48.7 & 0 & 0 & --\\
Random sets & 8 (7--10) & -- & -- & -- & -- & 0.51\\
\bottomrule
\end{tabular}
\end{table}

Table~\ref{tab:tmy} summarises the results. The median size of $\Chat$ was 8, $\Chat$ had more than five members in 85.2\% of station-months, and it contained the first-ranked candidate year in all but two station-months. MCB and the two rank projections contained $\Chat$ in every station-month and had median sizes between 11 and 16. The deepest candidate year, which $\Chat$ always contains, and the first-ranked candidate year coincided in only 48.7\% of station-months. The degree days over $\Chat$ ranged over a median of 20\% of the station-month mean, compared with 14\% for the TMY3 shortlist and 51\% for random sets of the same size. The centre set contained the whole TMY3 shortlist in 51.5\% of station-months.

Forty to fifty years of records do not determine the most central candidate year of a station-month, and the spread of degree days over $\Chat$ is an uncertainty in the main energy driver of a building simulation, which a single typical year does not show. In most station-months $\Chat$ has more than five members, and in about half of the station-months the TMY3 shortlist includes a candidate year that the data rule out, at the 5\% level, as the most central under the distance $\delta$. The members of $\Chat$ are much closer to one another in degree days than random candidate years are, so closeness under this distance carries over to the energy driver. We therefore suggest applying the secondary TMY3 checks to all members of $\Chat$ and reporting the range of simulated energy use over $\Chat$ together with the result for the chosen year.

\subsection{Gene trees}\label{sec:trees}

Phylogenomic studies estimate hundreds or thousands of gene trees, one for each locus, and the gene trees disagree with each other because of estimation error and real biological discordance. A central gene tree can be reported as a summary, and its features are then read as evidence about the species. We asked whether the data single out one gene tree as the most central, and which of its features are supported.

We analysed three collections: the 268 gene trees of seven species of Apicomplexa, single-celled parasites that include the malaria parasite \emph{Plasmodium}, together with the outgroup \emph{Tetrahymena}, distributed with the R package \texttt{kdetrees} \citep{weyenberg2014}; and the 329 intron and 3679 ultraconserved-element (UCE) gene trees that contain all 48 bird species of the Avian Phylogenomics Project \citep{jarvis2014}, available from GigaDB (\url{https://doi.org/10.5524/101041}). Each tree is represented by the vector of the lengths of its splits, the bipartitions of the taxa defined by its edges, with zero for a split the tree does not contain, and the distance between two trees is the Euclidean distance between these vectors, known as the branch-score distance. In each collection we computed $\Chat$. For the Apicomplexa trees, for which \citet{weyenberg2014}, using a different tree distance, identified twelve outlying genes, we also computed the outlier set $\widehat O_k$ for $k=27$, 10\% of the trees.

For the Apicomplexa trees we checked whether the trees in $\Chat$ have the most frequent topology, the branching pattern, ignoring branch lengths, shared by the largest number of gene trees. For the birds we checked which clades the trees in $\Chat$ contain. A clade consists of all the species descended from one common ancestor, and a tree contains it when one of its splits separates these species from all others. We considered three well-established clades, Palaeognathae, Galloanserae and Passeriformes, and the contested clade Columbea. Supplementary Section~S6 gives the details of the data and the clades.

\begin{table}[t]
\centering
\caption{Gene trees. P1--P3: Palaeognathae, Galloanserae and Passeriformes. $\widehat O_{27}$: number of trees flagged, with the number of outlying genes among them in brackets.}\label{tab:trees}
\footnotesize
\setlength{\tabcolsep}{4pt}
\begin{tabular}{lrccc}
\toprule
Collection & $n$ & $|\Chat|$ & Tree in $\Chat$ & $\widehat O_{27}$\\
\midrule
Apicomplexa & 268 & 1 & most frequent topology & 19 (12)\\
Introns & 329 & 1 & P1--P3, no Columbea & --\\
UCEs & 3679 & 1 & P1--P3, no Columbea & --\\
\bottomrule
\end{tabular}
\end{table}

Table~\ref{tab:trees} summarises the results. In each collection $\Chat$ contained a single tree, the deepest one, in line with Remark~\ref{rem:metric} for data of high intrinsic dimension. The central Apicomplexa tree has the most frequent topology in the collection. Both central bird trees contain the three well-established clades. Neither contains the contested clade Columbea. For the Apicomplexa trees, $\widehat O_{27}$ flagged 19 trees, including the twelve outlying genes of \citet{weyenberg2014}.

These results have two consequences for phylogenomic practice. First, one central tree can be reported in each collection, and by Section~\ref{sec:targets} its features, such as the clades above, are features of the most central tree under the distance used with probability at least 95\%. Second, the outlier sets can flag loci for inspection before the gene trees are combined into a species tree.

\section{Discussion}\label{sec:discussion}

We have developed simultaneous inference for rankings of random objects by the reference depth, using only a dissimilarity matrix. In general metric spaces, coverage of $\Chat$ is proved when $D>2(2-\gamma)$ and, under further local conditions, when $D<2$ (Remark~\ref{rem:metric}). Outside Euclidean space, the remaining range, including $D=2$, is open. In the location family the argument of Theorem~\ref{thm:1dlower} gives only a lower bound of constant order when $d\ge2$ (Supplementary Section~S3), so it remains open whether the logarithmic factors in Theorem~\ref{thm:highd}(i) are needed. On the line with $J\neq0$, the exact order of $|\Chat|$ is open (Section~\ref{sec:simgrowth}). Block multipliers are a natural way to extend the procedures to dependent objects, such as serially dependent candidate years.

For discrete objects such as tree topologies and rankings, distances often tie. With a tie counted as half a win in \eqref{eq:rhat}, the Hoeffding band of Section~\ref{sec:simband} keeps its coverage, as do its band centre set and band outlier sets, and the multiplier band extends to finitely supported distributions (Supplementary Section~S1). Coverage of $\Chat$ and $\widehat O_k$ under ties is open. The procedures of Section~\ref{sec:method} can be computed for any U-statistic depth, such as metric lens depth and metric spatial depth, and the multiplier band keeps its coverage when the kernel satisfies the conditions of Supplementary Section~S1; the centre-set theory of Sections~\ref{sec:1d} and \ref{sec:info} uses features specific to the reference depth. For metric spatial depth in Euclidean space, and for metric lens depth on the line, the influence function vanishes at the centre (Supplementary Section~S1), so the estimation error there has a different order, and centre sets for these depths remain open.

\section*{Acknowledgements}
The authors used Claude (Anthropic) for language editing and for checking proofs, code and numerical results. The authors verified all output and take full responsibility for the content.

\section*{Data availability}
All data analysed in this paper are publicly available. The hourly weather records are the ISD-Lite files of NOAA's Integrated Surface Database (\url{https://www.ncei.noaa.gov/pub/data/noaa/isd-lite/}). The Apicomplexa gene trees are distributed with the R package \texttt{kdetrees} \citep{weyenberg2014}, and the gene trees of the Avian Phylogenomics Project \citep{jarvis2014} are available from GigaDB (\url{https://doi.org/10.5524/101041}). The Python package \pkgname\ and the code that reproduces every number, table and figure accompany the submission and will be made publicly available on publication.

\bibliographystyle{agsm}
\bibliography{ref_jrssb}

@article{Yeon2025JRSSB,
    author = {Yeon, Hyemin and Dai, Xiongtao and Lopez-Pintado, Sara},
    title = {Regularized halfspace depth for functional data},
    journal = {Journal of the Royal Statistical Society Series B: Statistical Methodology},
    volume = {87},
    number = {5},
    pages = {1553-1575},
    year = {2025},
    issn = {1369-7412},
    doi = {10.1093/jrsssb/qkaf030},
}

@article{kleindessner2017lensdepth,
  title={Lens depth function and {$k$}-relative neighborhood graph: versatile tools for ordinal data analysis},
  author={Kleindessner, Matth{\"a}us and von Luxburg, Ulrike},
  journal={Journal of Machine Learning Research},
  volume={18},
  number={58},
  pages={1--52},
  year={2017}
}

@article{liu2011lens,
  title={Lens data depth and median},
  author={Liu, Zhenyu and Modarres, Reza},
  journal={Journal of Nonparametric Statistics},
  volume={23},
  number={4},
  pages={1063--1074},
  year={2011},
  publisher={Taylor \& Francis}
}

@inproceedings{Tukey1975,
  title={Mathematics and the picturing of data},
  author={Tukey, John W},
  booktitle={Proceedings of the International Congress of Mathematicians, Vancouver, 1974},
  volume={2},
  pages={523--531},
  year={1975},
  publisher={Canadian Mathematical Congress}
}

@article{deb2021multivariate,
  title={Multivariate rank-based distribution-free nonparametric testing using measure transportation},
  author={Deb, Nabarun and Sen, Bodhisattva},
  journal={Journal of the American Statistical Association},
  volume={118},
  number={541},
  pages={192--207},
  year={2023},
  publisher={Taylor \& Francis}
}

@article{Zuo2003,
  title={Projection-based depth functions and associated medians},
  author={Zuo, Yijun},
  journal={The Annals of Statistics},
  volume={31},
  number={5},
  pages={1460--1490},
  year={2003},
  publisher={Institute of Mathematical Statistics}
}

@article{Liu1990,
author = {Regina Y. Liu},
title = {On a notion of data depth based on random simplices},
volume = {18},
year = {1990},
journal = {The Annals of Statistics},
number = {1},
publisher = {Institute of Mathematical Statistics},
pages = {405--414},
doi = {10.1214/aos/1176347507},
}

@InProceedings{zhou2023new,
  title = 	 {A new ranking scheme for modern data and its application to two-sample hypothesis testing},
  author =       {Zhou, Doudou and Chen, Hao},
  booktitle = 	 {Proceedings of Thirty Sixth Conference on Learning Theory},
  pages = 	 {3615--3668},
  year = 	 {2023},
  volume = 	 {195},
  series = 	 {Proceedings of Machine Learning Research},
  publisher =    {PMLR},
}

@article{pan2018ball,
  title={Ball divergence: nonparametric two sample test},
  author={Pan, Wenliang and Tian, Yuan and Wang, Xueqin and Zhang, Heping},
  journal={The Annals of Statistics},
  volume={46},
  number={3},
  pages={1109--1137},
  year={2018}
}

@article{hallin2002optimal,
  title={Optimal tests for multivariate location based on interdirections and pseudo-{M}ahalanobis ranks},
  author={Hallin, Marc and Paindaveine, Davy},
  journal={The Annals of Statistics},
  volume={30},
  number={4},
  pages={1103--1133},
  year={2002},
  publisher={Institute of Mathematical Statistics}
}

@article{zuo2000general,
  title={General notions of statistical depth function},
  author={Zuo, Yijun and Serfling, Robert},
  journal={The Annals of Statistics},
  volume={28},
  number={2},
  pages={461--482},
  year={2000},
  publisher={JSTOR}
}

@article{liu1993quality,
  title={A quality index based on data depth and multivariate rank tests},
  author={Liu, Regina Y and Singh, Kesar},
  journal={Journal of the American Statistical Association},
  volume={88},
  number={421},
  pages={252--260},
  year={1993},
  publisher={Taylor \& Francis}
}

@book{oja2010multivariate,
  title={Multivariate Nonparametric Methods with {R}: An Approach Based on Spatial Signs and Ranks},
  author={Oja, Hannu},
  series={Lecture Notes in Statistics},
  volume={199},
  year={2010},
  publisher={Springer},
  address={New York}
}

@article{chaudhuri1996geometric,
  title={On a geometric notion of quantiles for multivariate data},
  author={Chaudhuri, Probal},
  journal={Journal of the American Statistical Association},
  volume={91},
  number={434},
  pages={862--872},
  year={1996},
  publisher={Taylor \& Francis}
}

@article{puri1966class,
  title={On a class of multivariate multisample rank-order tests},
  author={Puri, Madan Lal and Sen, Pranab Kumar},
  journal={Sankhy{\=a}: The Indian Journal of Statistics, Series A},
  volume={28},
  number={4},
  pages={353--376},
  year={1966},
  publisher={JSTOR}
}

@article{hallin1995multivariate,
  title={A multivariate {W}ald--{W}olfowitz rank test against serial dependence},
  author={Hallin, Marc and Puri, Madan L},
  journal={Canadian Journal of Statistics},
  volume={23},
  number={1},
  pages={55--65},
  year={1995},
  publisher={Wiley Online Library}
}

@article{bradley1952rank,
  title={Rank analysis of incomplete block designs: {I}. {T}he method of paired comparisons},
  author={Bradley, Ralph Allan and Terry, Milton E},
  journal={Biometrika},
  volume={39},
  number={3--4},
  pages={324--345},
  year={1952},
  publisher={JSTOR}
}

@misc{liu2022quantiles,
  title  = {Quantiles, Ranks and Signs in Metric Spaces},
  author = {Liu, Hang and Wang, Xueqin and Zhu, Jin and Zhang, Heping},
  year   = {2022},
  note   = {arXiv:2209.04090},
}

@article{wang2024mdf,
  title   = {Nonparametric Statistical Inference via Metric Distribution Function in Metric Spaces},
  author  = {Wang, Xueqin and Zhu, Jin and Pan, Wenliang and Zhu, Junhao and Zhang, Heping},
  journal = {Journal of the American Statistical Association},
  volume  = {119},
  number  = {548},
  pages   = {2772--2784},
  year    = {2024},
  doi     = {10.1080/01621459.2023.2277417},
}

@article{geenens2023statistical,
  title   = {Statistical depth in abstract metric spaces},
  author  = {Geenens, Gery and Nieto-Reyes, Alicia and Francisci, Giacomo},
  journal = {Statistics and Computing},
  volume  = {33},
  number  = {2},
  pages   = {46},
  year    = {2023},
}

@article{dai2023tukey,
  title   = {Tukey's depth for object data},
  author  = {Dai, Xiongtao and Lopez-Pintado, Sara},
  journal = {Journal of the American Statistical Association},
  volume  = {118},
  number  = {543},
  pages   = {1760--1772},
  year    = {2023},
}

@article{virta2023,
  title   = {Spatial depth for data in metric spaces},
  author  = {Virta, Joni},
  journal = {Scandinavian Journal of Statistics},
  volume  = {53},
  number  = {2},
  pages   = {684--711},
  year    = {2026},
  doi     = {10.1111/sjos.70054},
}

@article{mogstad2024inference,
  title   = {Inference for ranks with applications to mobility across neighbourhoods and academic achievement across countries},
  author  = {Mogstad, Magne and Romano, Joseph P. and Shaikh, Azeem M. and Wilhelm, Daniel},
  journal = {The Review of Economic Studies},
  volume  = {91},
  number  = {1},
  pages   = {476--518},
  year    = {2024},
}

@article{hsu1981,
  title   = {Simultaneous confidence intervals for all distances from the ``best''},
  author  = {Hsu, Jason C.},
  journal = {The Annals of Statistics},
  volume  = {9},
  number  = {5},
  pages   = {1026--1034},
  year    = {1981},
}

@article{gupta1965,
  title   = {On some multiple decision (selection and ranking) rules},
  author  = {Gupta, Shanti S.},
  journal = {Technometrics},
  volume  = {7},
  number  = {2},
  pages   = {225--245},
  year    = {1965},
}

@article{hansen2011mcs,
  title   = {The model confidence set},
  author  = {Hansen, Peter R. and Lunde, Asger and Nason, James M.},
  journal = {Econometrica},
  volume  = {79},
  number  = {2},
  pages   = {453--497},
  year    = {2011},
}

@article{andrewssoares2010,
  title   = {Inference for parameters defined by moment inequalities using generalized moment selection},
  author  = {Andrews, Donald W. K. and Soares, Gustavo},
  journal = {Econometrica},
  volume  = {78},
  number  = {1},
  pages   = {119--157},
  year    = {2010},
}

@article{gao2023uncertainty,
  title   = {Uncertainty quantification in the {B}radley--{T}erry--{L}uce model},
  author  = {Gao, Chao and Shen, Yandi and Zhang, Anderson Y.},
  journal = {Information and Inference: A Journal of the IMA},
  volume  = {12},
  number  = {2},
  pages   = {1073--1140},
  year    = {2023},
}

@article{bates2023testing,
  title   = {Testing for outliers with conformal p-values},
  author  = {Bates, Stephen and Cand{\`e}s, Emmanuel and Lei, Lihua and Romano, Yaniv and Sesia, Matteo},
  journal = {The Annals of Statistics},
  volume  = {51},
  number  = {1},
  pages   = {149--178},
  year    = {2023},
}

@book{vdvw1996,
  title     = {Weak Convergence and Empirical Processes: With Applications to Statistics},
  author    = {van der Vaart, Aad W. and Wellner, Jon A.},
  publisher = {Springer},
  address   = {New York},
  year      = {1996},
}

@incollection{houdre2003,
  title     = {Exponential inequalities, with constants, for {U}-statistics of order two},
  author    = {Houdr{\'e}, Christian and Reynaud-Bouret, Patricia},
  booktitle = {Stochastic Inequalities and Applications},
  editor    = {Gin{\'e}, Evarist and Houdr{\'e}, Christian and Nualart, David},
  series    = {Progress in Probability},
  volume    = {56},
  pages     = {55--69},
  publisher = {Birkh{\"a}user},
  address   = {Basel},
  year      = {2003},
}

@techreport{wilcox2008,
  title       = {Users Manual for {TMY3} Data Sets},
  author      = {Wilcox, Stephen and Marion, William},
  institution = {National Renewable Energy Laboratory},
  number      = {NREL/TP-581-43156},
  year        = {2008},
}

@techreport{hall1978tmy,
  title       = {Generation of Typical Meteorological Years for 26 {SOLMET} Stations},
  author      = {Hall, I. J. and Prairie, R. R. and Anderson, H. E. and Boes, E. C.},
  institution = {Sandia Laboratories},
  number      = {SAND78-1601},
  year        = {1978},
}

@article{jarvis2014,
  title   = {Whole-genome analyses resolve early branches in the tree of life of modern birds},
  author  = {Jarvis, Erich D. and others},
  journal = {Science},
  volume  = {346},
  number  = {6215},
  pages   = {1320--1331},
  year    = {2014},
}

@article{weyenberg2014,
  title   = {{kdetrees}: non-parametric estimation of phylogenetic tree distributions},
  author  = {Weyenberg, Grady and Huggins, Peter M. and Schardl, Christopher L. and Howe, Daniel K. and Yoshida, Ruriko},
  journal = {Bioinformatics},
  volume  = {30},
  number  = {16},
  pages   = {2280--2287},
  year    = {2014},
}

@article{nye2017pca,
  title   = {Principal component analysis and the locus of the {F}r{\'e}chet mean in the space of phylogenetic trees},
  author  = {Nye, Tom M. W. and Tang, Xiaoxian and Weyenberg, Grady and Yoshida, Ruriko},
  journal = {Biometrika},
  volume  = {104},
  number  = {4},
  pages   = {901--922},
  year    = {2017},
}

@misc{lyu2026core,
  title  = {Preference-based Centrality and Ranking in General Metric Spaces},
  author = {Lyu, Lingfeng and Zhou, Doudou},
  year   = {2026},
  note   = {arXiv:2601.18412},
}

@article{dubey2024metric,
  title   = {Metric statistics: Exploration and inference for random objects with distance profiles},
  author  = {Dubey, Paromita and Chen, Yaqing and M{\"u}ller, Hans-Georg},
  journal = {The Annals of Statistics},
  volume  = {52},
  number  = {2},
  pages   = {757--792},
  year    = {2024},
  doi     = {10.1214/24-AOS2368},
}

@article{zamanifarizhandi2024oja,
  title   = {Metric {O}ja depth, new statistical tool for estimating the most central objects},
  author  = {Zamanifarizhandi, Vida and Virta, Joni},
  journal = {Computational Statistics \& Data Analysis},
  volume  = {224},
  pages   = {108447},
  year    = {2026},
  doi     = {10.1016/j.csda.2026.108447},
}

@article{lei2020cvc,
  title   = {Cross-validation with confidence},
  author  = {Lei, Jing},
  journal = {Journal of the American Statistical Association},
  volume  = {115},
  number  = {532},
  pages   = {1978--1997},
  year    = {2020},
}

@misc{kim2025argmin,
  title  = {Locally minimax optimal confidence sets for the best model},
  author = {Kim, Ilmun and Ramdas, Aaditya},
  year   = {2025},
  note   = {arXiv:2503.21639},
}

@article{davies1993identification,
  title   = {The identification of multiple outliers},
  author  = {Davies, Laurie and Gather, Ursula},
  journal = {Journal of the American Statistical Association},
  volume  = {88},
  number  = {423},
  pages   = {782--792},
  year    = {1993},
}

@article{dang2010nonparametric,
  title   = {Nonparametric depth-based multivariate outlier identifiers, and masking robustness properties},
  author  = {Dang, Xin and Serfling, Robert},
  journal = {Journal of Statistical Planning and Inference},
  volume  = {140},
  number  = {1},
  pages   = {198--213},
  year    = {2010},
}

@article{clemencon2023affine,
  author = {Cl{\'e}men{\c c}on, Stephan and Mozharovskyi, Pavlo and Staerman, Guillaume},
  title = {Affine invariant integrated rank-weighted statistical depth: properties and finite sample analysis},
  year = {2023},
  journal = {Electronic Journal of Statistics},
  volume = {17},
  number = {2},
  pages = {3854--3892},
  doi = {10.1214/23-ejs2189},
}

@article{ramsay2019integrated,
  author = {Ramsay, Kelly and Durocher, St{\'e}phane and Leblanc, Alexandre},
  title = {Integrated rank-weighted depth},
  year = {2019},
  journal = {Journal of Multivariate Analysis},
  volume = {173},
  pages = {51--69},
  doi = {10.1016/j.jmva.2019.02.001},
}

@misc{jentsch2026resampling,
  author = {Jentsch, Carsten and Nagy, Stanislav and Wendler, Martin},
  title = {Resampling simplicial depth},
  year = {2026},
  eprint = {2608.11131},
  archivePrefix = {arXiv},
  primaryClass = {stat.ME},
  doi = {10.48550/arXiv.2608.11131},
  note = {arXiv:\allowbreak 2608.11131},
}

@article{masse2004asymptotics,
  author = {Mass{\'e}, Jean-Claude},
  title = {Asymptotics for the {T}ukey depth process, with an application to a multivariate trimmed mean},
  year = {2004},
  journal = {Bernoulli},
  volume = {10},
  number = {3},
  pages = {397--419},
  doi = {10.3150/bj/1089206404},
}

@article{dumbgen1992limit,
  author = {D{\"u}mbgen, Lutz},
  title = {Limit theorems for the simplicial depth},
  year = {1992},
  journal = {Statistics \& Probability Letters},
  volume = {14},
  number = {2},
  pages = {119--128},
  doi = {10.1016/0167-7152(92)90075-g},
}

@article{yehsingh1997balanced,
  author = {Yeh, Arthur B. and Singh, Kesar},
  title = {Balanced Confidence Regions Based on {T}ukey’s Depth and the Bootstrap},
  year = {1997},
  journal = {Journal of the Royal Statistical Society Series B: Statistical Methodology},
  volume = {59},
  number = {3},
  pages = {639--652},
  doi = {10.1111/1467-9868.00088},
}

@article{liusingh1997limiting,
  author = {Liu, Regina Y. and Singh, Kesar},
  title = {Notions of Limiting P Values Based on Data Depth and Bootstrap},
  year = {1997},
  journal = {Journal of the American Statistical Association},
  volume = {92},
  number = {437},
  pages = {266--277},
  doi = {10.1080/01621459.1997.10473624},
}

@article{narisetty2016extremal,
  author = {Narisetty, Naveen N. and Nair, Vijayan N.},
  title = {Extremal Depth for Functional Data and Applications},
  year = {2016},
  journal = {Journal of the American Statistical Association},
  volume = {111},
  number = {516},
  pages = {1705--1714},
  doi = {10.1080/01621459.2015.1110033},
}

@article{arcones1992bootstrap,
  author = {Arcones, Miguel A. and Gin{\'e}, Evarist},
  title = {On the Bootstrap of {$U$} and {$V$} Statistics},
  year = {1992},
  journal = {The Annals of Statistics},
  volume = {20},
  number = {2},
  pages = {655--674},
  doi = {10.1214/aos/1176348650},
}

@article{kimpark2026conformal,
  author = {Kim, Hyunsung and Park, Junyong},
  title = {Conformal outlier detection for multivariate functional data},
  year = {2026},
  journal = {Computational Statistics},
  volume = {41},
  number = {4},
  eid = {88},
  pages = {88},
  doi = {10.1007/s00180-026-01763-1},
}

@article{febrero2008outliers,
  author = {Febrero, Manuel and Galeano, Pedro and Gonz{\'a}lez-Manteiga, Wenceslao},
  title = {Outlier detection in functional data by depth measures, with application to identify abnormal {NO$_x$} levels},
  year = {2008},
  journal = {Environmetrics},
  volume = {19},
  number = {4},
  pages = {331--345},
  doi = {10.1002/env.878},
}

@article{hoeffding1963,
  author = {Hoeffding, Wassily},
  title = {Probability Inequalities for Sums of Bounded Random Variables},
  journal = {Journal of the American Statistical Association},
  year = {1963},
  volume = {58},
  number = {301},
  pages = {13--30},
  doi = {10.1080/01621459.1963.10500830}
}

\end{document}